\documentclass[journal,onecolumn]{IEEEtran}

\usepackage{commath}
\usepackage{amsmath,amsfonts,amssymb,amsthm}
\usepackage{mathtools}
\usepackage{commath}
\usepackage[sc,osf]{mathpazo}
\usepackage[square,numbers]{natbib}
\usepackage{graphicx}
\usepackage{grffile}

{}
{}

\theoremstyle{remark}
\newtheorem{remark}{Remark}

\begin{document}
\title{Fast Zonotope-Tube-based LPV-MPC for Autonomous Vehicles}

\author{Eugenio~Alcal\'a, Vicen\c{c}~Puig, Joseba~Quevedo and Olivier~Sename
\thanks{A. Alcal\'a, V. Puig and J. Quevedo are with the Department of Automatic Control, Polytechnic University of Catalonia, Barcelona, Spain e-mail: vicenc.puig@upc.edu.}
\thanks{O. Sename is with the Control System Department at GIPSA-lab, Grenoble, France}}


\maketitle

\begin{abstract}
In this paper, we present an effective online tube-based model predictive control (T-MPC) solution for autonomous driving that aims at improving the computational load while ensuring robust stability and performance in fast and disturbed scenarios.
We focus on reformulating the non-linear original problem into a pseudo-linear problem by transforming the non-linear vehicle equations to be expressed in a Linear Parameter Varying (LPV) form. An scheme composed by a nominal controller and a corrective local controller is propossed.
First, the local controller is designed as a polytopic LPV-H$_{\infty}$ controller able to reject external disturbances. Moreover, a finite number of accurate reachable sets, also called tube, are computed online using zonotopes taking into account the system dynamics, the local controller and the diturbance-uncertainty bounds considered. 
Second, the nominal controller is designed as an MPC where the LPV vehicle model is used to speed up the computational time while keeping accurate vehicle representation. 
Employing reachability theory with zonotopes, the MPC changes online its state and input constraints to ensure robust feasibility and stability under exhogenous disturbances.
Finally, we test the presented scheme and compare the local controller performance against the LQR design as state of the art approach.
We demonstrate its effectiveness in a disturbed fast driving scenario being able to reject strong exogenous disturbances and fulfilling imposed constraints at a very reduced computational cost.
\end{abstract}


\maketitle

\section{Introduction}
In last recent years, the number of vehicles on the roads has grown significantly and subsequently the risk of car accidents.
In a near future, when autonomous vehicles are finally in the streets, we will expect them to handle the most challenging situations that humans handle nowadays. 
They will have to deal with the complete net of transportation, i.e. vehicles, traffic rules, pedestrians etc, but also with extreme weather situations. 
Most of these cases can be either forecasted or approximated by rules since they follow known physical behaviours in the weather case or traffic rules in the case of vehicles and pedetrians. 
However, sometimes this may not happen as expected and the vehicle is suddenly running into extreme situations such as for example very windy situation on highways.
Addressing these situations is what control engineering refers to as robustness: the ability of the controller to handle unexpected situations such as internal variations in the system or in the external environment affecting the system.

\noindent A large variety of control strategies have been studied to address the robustness in control of systems.
All these methods pursue the same objective: ensure asymptotic stability, robustness and performance \citep{Weinmann, Kouvaritakis}.

\noindent Model Predictive Control (MPC) is an effective control strategy that allows to deal with constrained problems and multiple-input multiple-output systems.
However, dealing with uncertainty or disturbances is something that conventional MPC algorithms do not handle and then, robust MPC (RMPC) formulations have to be considered where the design is done by means of robustifying the constraints \citep{RMPC_Uncertainty_Sets}. 
In \cite{Mayne_MPC_review}, the author presents a review on current MPC formulations with their limitations and future development directions.

\noindent During the last years, two differentiated and consolidated approaches for robust MPC have been addressed: Min-max MPC and Tube-based MPC (T-MPC).
On the one hand, the min-max or worst-case problem aims to find the optimal solution based on minimizing the maximum value of the cost function. In \citep{min_max_MPC}, authors present a robust self-triggered min-max MPC approach for constrained non-linear systems with both parameter uncertainties and disturbances.
On the other hand, T-MPC is based on computing a region around the nominal prediction that ensures the state of the system to remain inside under any possible uncertainty and disturbance \citep{tube_MPC}. \\

\begin{figure}[h!]
    	\centering
    	\includegraphics[width=85mm]{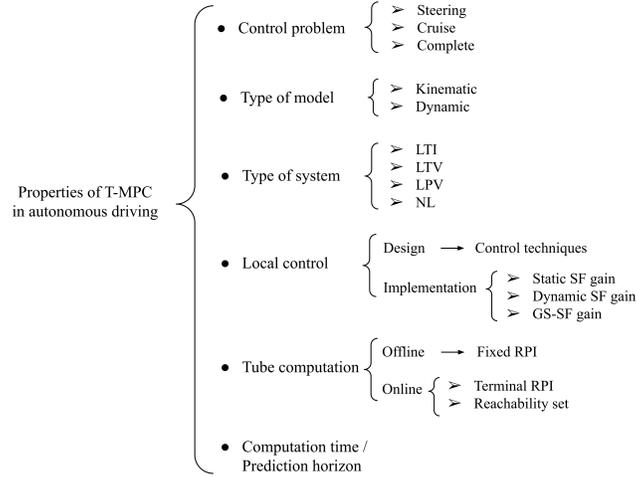}
    	\caption{ Diagram of different characteristics involved in the design of T-MPC technique in the self-driving car field}
    \label{fig:T-MPC_classification}
\end{figure}

\noindent Our work is mostly inspired by the T-MPC technique. This strategy has been widely employed in the mobile robotics field \citep{Gonzalez_R_2011, Kayacan_2015, Sun_2017, Sun_2018}. 
However, from a self-driving car perspective we do not find many references in the literature.
In Figure \ref{fig:T-MPC_classification}, we show a diagram made for classifying the T-MPC technique applyied to autonomous driving as function of some involved design characteristics.
Furthermore, we present in Table \ref{table:T_MPC_Classification} a review of those works dealing with T-MPC in autonomous driving of cars subject to the properties presented in Figure \ref{fig:T-MPC_classification}.

\begin{table*}[htbp]\label{table:T_MPC_Classification}
\caption{Classification of T-MPC technique in autonomous driving field. Some used acronyms: SF := state feedback, LQR := linear quadratic regulator, LTI := linear time invariant, RPI := robust positively invariant, GS := gain scheduling}
{\begin{tabular*}{40pc}{@{\extracolsep{\fill}}llll@{}}
\hline
Work & Sakhdari, B., et al. (2017)\citep{Sakhdari_B_2017} & Rathai, K., et al. (2017)\citep{Rathai_K} & Bujarbaruah, M., et al. (2018)\citep{Bujarbaruah_M} \\
\hline
Control problem & Cruise control & Steering control & Steering control  \\
Type of model & Long. dynamics & Lateral dynamics & Lateral. dynamics \\
Type of system & LTI & LPV & LTI \\
Local control / Design & LQR & Polytopic LQR & LQR  \\
Local control / Implement. & Static SF gain & Static SF gain & Static SF gain  \\
Tube computation & Polytopic aligned-box & minimal RPI & online terminal RPI  \\
Computational time / Horizon & No info & 25 ms / 7 steps & 100 ms / 6 steps  \\
\hline
Work & Sakhdari, B., et al. (2018)\citep{Sakhdari_B_2018} & Mata, S., et al. (2019)\citep{Mata_S} & Our approach \\
\hline
Control problem & Cruise control & Steering control & Complete control \\
Type of model & Long. dynamics & Lateral. dynamics & Complete dynamics \\
Type of system & LTI & LTI & LPV \\
Local control / Design & Linear & No info & Polytopic H$_{\infty}$ \\
Local control / Implement. & Static SF gain & No info & GS-SF gain \\
Tube computation & offline fixed invariant& online RPI & online adaptive zonotopic \\
Computational time / Horizon & $<$ 1 ms / 10 steps & 100 ms / 15 steps & 33 ms / 5 steps \\ 
\hline
\end{tabular*}}{}
\end{table*}

\noindent In this paper, we present a robust T-MPC approach faster than the state of the art strategies being able to reject large exogenous disturbances. 
This optimal algorithm uses a Linear Parameter Varying (LPV) vehicle model for simulating future behaviour. The principal concept behind the LPV modeling approach is that the non-linear model representation can be expressed as a combination of linear models that depend on some scheduling variables without using linearization \citep{Olivier_LPV}.
Furthermore, the introduction of zonotope-based operations to compute reachable sets allow to make the algorithm faster and more accurate. \\

\noindent We summarize the innovative points with respect to the state of the art as follows:
\begin{itemize}
	\item Using zonotope theory we are able to reduce the computational cost of basic operations, i.e. Minkowsky sum and difference, in comparison with standard polytopes-based operations.
	
	\item The use of zonotope-based calculations allows to bound more tightly the tube, hence obtaining a less conservative result and more accurate result.
	
	\item Using $H_{\infty}$ control design to obtain a gain scheduling polytopic LPV local controller allows to reject large exogenous disturbances acting over the vehicle. Current T-MPC techniques in the state of the art are using LQR technique.
	
	\item Currently, most of the works based on robust MPC design use a local controller than runs at the same frequency than the nominal controller (MPC). In this work, we propose a faster loop to achieve a faster and better performance of the control scheme.

\end{itemize} 

\noindent The paper is structured as follows: Section 2 presents the problem statement.
Section 3 addresses the core of this work: the online T-LPV-MPC using zonotopes algorithm.
In Section 4, we present the main results and a proper discussion.
Finally, last section presents the conlusions of the work.


\section{General problem statement}

\begin{figure}[h!]
    	\centering
    	\includegraphics[width=85mm]{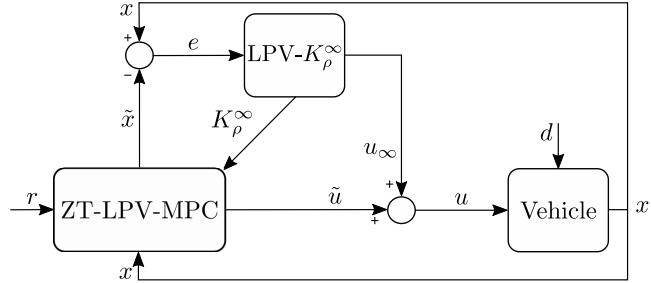}
    	\caption{ Robust control scheme composed of a nominal controller (ZT-LPV-MPC) and a local corrective controller (LPV-$K_{\zeta}^{\infty}$)}
    \label{fig:Robust_MPC_Kinf_escheme}
\end{figure}

This paper addresses the problem of designing an online T-MPC for controlling a simulated vehicle plant (see Figure \ref{fig:Robust_MPC_Kinf_escheme}) formulated as the following non-linear system
\begin{equation}
	\label{eq:NL-syst}
	x^+ = f(x,u) + e \ ,
\end{equation}
where $x \in \mathbb{R}^n$ is the state vector, $u \in \mathbb{R}^m$ the input vector and $f(x,u)$ represents the non-linear map obtained after modeling the physics of the real system.
Vector $e \in \mathbb{R}^n$ contains all the unmodeled physics of the real plant and exogenous disturbances acting over it. 
Note that, in the following we will introduce $d$ as the disturbance vector and this is contained in $e$.
Besides, note that in this article the notation $x^+$ is used for the successor of vector $x$, i.e. $x = x(k)$ and $x^+ = x(k+1)$. \\

\noindent At this point, the following uncertain, LPV, discrete-time system is formulated
\begin{equation}
	\label{eq:linear-syst-formulation}
	x^{+} = A_{\zeta} x + B_{\zeta} u + w \ ,
\end{equation}
where $A_{\zeta} \in \mathbb{R}^{n \times n}$ and $B_{\zeta} \in \mathbb{R}^{n \times m}$ are the LPV state space matrices which depend on the varying scheduling vector $\zeta \in \mathbb{R}^N$, being $N$ the number of scheduling variables. 

\begin{remark}
	The system $x^{+} = A_{\zeta} x + B_{\zeta} u$ in \eqref{eq:linear-syst-formulation} is an exact realization of $x^+ = f(x,u)$ in \eqref{eq:NL-syst} using the non-linear embedding approach inside the considered polytopic region and the scheduling vector $\zeta$ is known at each sampling time being a combination of system states and inputs.
\end{remark}

\noindent The state, control and disturbance vectors are bounded as
\begin{equation}
	x \in X \ , \ \ u \in U \ , \ \ w \in W \ ,	 
\end{equation}
where $X \subseteq \mathbb{R}^n$, $U \subseteq \mathbb{R}^m$ and $W \subseteq \mathbb{R}^n$. \\

\noindent To achieve the tracking and robust control purposes, two problems are handled:
\begin{itemize}
	\item \textbf{Reference tracking control problem}. The LPV-MPC strategy deals with the following disturbance-free system for tracking the dynamic references while handling system constraints
\begin{equation}
	\label{eq:nominal-model}
	\tilde{x}^{+} = A_{\zeta} \tilde{x} + B_{\zeta} \tilde{u} ,
\end{equation}
where the state ($\tilde{x} \in \mathbb{R}^n$) and optimal control ($\tilde{u} \in \mathbb{R}^m$) vectors are bounded as
\begin{equation}
	\tilde{x} \in \tilde{X} \ , \ \ \tilde{u} \in \tilde{U} \ ,	 
\end{equation}
where $\tilde{X} \subseteq \mathbb{R}^n$ and $\tilde{U} \subseteq \mathbb{R}^m$.
System \eqref{eq:nominal-model} will be refered as nominal model throughout the work.

	\item \textbf{Robust control problem}. 
		The main idea is to compensate the mismatch between the states of \eqref{eq:NL-syst} and the nominal state vectors \eqref{eq:nominal-model}.
		This difference is computed as
		\begin{equation}
			e = x - \tilde{x} \ ,
		\end{equation}
		where $e$ is the error state.
		In order to minimize such a mismatch, the following control law is considered
		\begin{equation}
			\begin{aligned}
			u = \tilde{u} + u_{\infty}  \\
			u_{\infty} = K_{\zeta}^{\infty} e \ ,
			\end{aligned}
		\end{equation}
		where $u_{\infty} \in \mathbb{R}^m$ is the corrective action and $K_{\zeta}^{\infty} \in \mathbb{R}^{m \times n}$ is the state feedback gain computed online as a gain scheduling controller using the $H_{\infty}$-LMI-based problem for the design.
		
		\noindent Finally, the closed loop error dynamics are defined as
		\begin{equation}
		\label{eq:error_dynamics}
			e^+ = x^+ - \tilde{x}^+ = ( A_{\zeta} + B_{\zeta} K_{\zeta}^{\infty} ) e + w \ .
		\end{equation}
	\end{itemize}

\section{Vehicle modeling}    
An LPV system is a dynamical system of finite dimension whose state space matrices are fixed functions of a vector of measurable scheduling variables.

\noindent Obtaining the LPV formulation of a non-linear system may be sometimes a non trivial task. 
Particularly, trying to obtain the LPV representation presented in \eqref{eq:linear-syst-formulation} may result on many different options and not all of them with the same quality representation.

\noindent Then, the non-linear equations considered in this work defining the behaviour of the vehicle are
\begin{equation}
    \label{eq:NL_model_for_LPV}
    \begin{aligned}
       	& \dot v_x = a_r + \frac{- F_{yf} \sin{\delta} - F_{df}}{m} + \omega v_y  \\
        	& \dot v_y =  \frac{ F_{yf} \cos{\delta} + F_{yr}}{m} - \omega v_x \\
		& \dot \omega = \frac{ F_{yf} l_f \cos{\delta} - F_{yr} l_r}{I} \\
		& \dot x = v_x \\
		& \dot \theta = \omega \\
    		& \alpha_f = \delta -  \frac{v_y}{v_x} + \frac{l_f \omega}{v_x}  \\
    		& \alpha_r = - \frac{v_y}{v_x} - \frac{l_r \omega}{v_x} \\        
		&  F_{yf} = C_f(\alpha_f) \alpha_f \\
		&  F_{yr} = C_r(\alpha_r) \alpha_r \\
		& F_{df} = \mu m g + \frac{1}{2} \rho C_{dAf} v_x^2 \ . \        
    \end{aligned}
\end{equation}
\begin{figure}[h!]
    	\centering
    	\includegraphics[width=75mm]{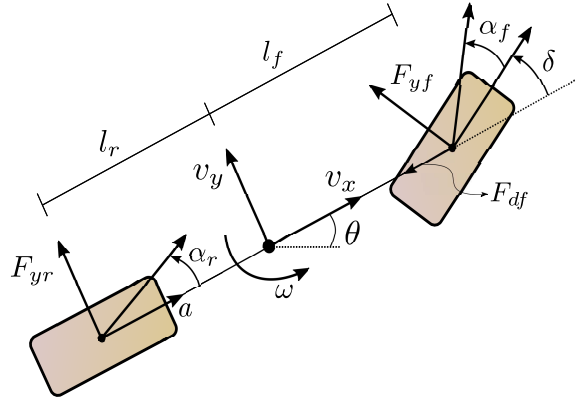}
    	\caption{ Illustration of the bicycle vehicle model}
    \label{fig:bicycle_representation}
\end{figure}  

\noindent where the dynamic vehicle variables $v_x$, $v_y$ and $\omega$ represent the body frame velocities, i.e. linear in $x$, linear in $y$ and angular velocities, respectively. 
Variables $x$ and $\theta$ are the integral with respect of time of $v_x$ and $\omega$, respectively.
The control variables $\delta$ and $a$ are the steering angle at the front wheels and the longitudinal acceleration vector on the rear wheels, respectively.
$F_{yf}$ and $F_{yr}$ are the lateral forces produced in front and rear tires, respectively.
Both $C_{f}(\alpha_f)$ and $C_{r}(\alpha_r)$ represent the front and rear tire stiffness coefficient non-linear functions, respectively.
Front and rear slip angles are represented as $\alpha_f$ and $\alpha_r$, respectively.
$m$ and $I$ represent the vehicle mass and inertia and $l_f$ and $l_r$ are the distances from the vehicle center of mass to the front and rear wheel axes, respectively. $\mu$, $\rho$ and $g$ are the friction coefficient, the air density and the gravity values, respectively.
$C_{dAf}$ is the product of drag coefficient and vehicle frontal cross sectional area.
All the dynamic vehicle parameters are properly defined in Table \ref{table:vehicle_parameters} and showed in Figure \ref{fig:bicycle_representation}.    
   
\noindent Then, denoting the state and control vectors, respectively, as

    \begin{subequations}
    \label{eq:LPV_model}
        \begin{equation}
            \label{eq:dynamic_state_space_vectors}
            x =
            \left[\begin{array}{c}
                v_x \\
                v_y  \\
                \omega  \\
                x \\
                \theta \\
            \end{array}\right] , \
            u =
        	\left[\begin{array}{c}                
                \delta \\
                 a\\
            \end{array}\right] ,
        \end{equation}

\noindent the non-linear model \eqref{eq:NL_model_for_LPV} is transformed into the discrete LPV representation \eqref{eq:nominal-model} by embedding the non-linearities within varying parameters. The matrices are linear dependent on the following scheduling variables 
		\begin{equation}
		\label{eq:scheduling_vector}
			\zeta := \left[ v_x, v_y, \delta \right]
		\end{equation}
and such a scheduling variables may vary in a non-linear way. 
In addition, the non-linear functions $C_{f}(\alpha_f)$ and $C_{r}(\alpha_r)$ are also formulated as an LPV representation and presented in Appendix B.	

\noindent Then, the continuous-time LPV matrices, i.e. $A_{\zeta}$ and $B_{\zeta}$, are formulated as
		\begin{equation}
        	A_{\zeta} =	\left[\begin{array}{ccccc}
                    A_{11}  & A_{12}    & A_{13} & 0 & 0   \\
                    0       & A_{22}    & A_{23} & 0 & 0   \\
                    0       & A_{32}    & A_{33} & 0 & 0   \\
                    -1    	 &   0      & 0      & 0 & 0   \\
                    0       &   0       & -1    & 0 & 0   \\      
        	\end{array}\right] \ ,
 	   	\end{equation}
 	   	and
 	   	\begin{equation}
 	   	\label{eq:LPV_B}
        	B_{\zeta} = \left[\begin{array}{cc}
                    -\frac{1}{m} \sin{\delta} C_f & 1 \\
                    \frac{1}{m} \cos{\delta} C_f & 0 \\
                    \frac{1}{I} \cos{\delta} C_f l_f  & 0 \\
                    0 & 0 \\
                    0 & 0 \\
        	\end{array}\right] \ ,
		\end{equation}
		\noindent being 
		\begin{equation}	
		\label{eq:final_LPV_AB_elements}
		\begin{array}{c}		
        A_{11} = \frac{-\mu g}{v_x} - \frac{\rho C_{dA} v_x}{2m}\\
        A_{12} = \frac{C_f \sin{\delta} }{m v_x} \\     
        A_{13} = \frac{C_f l_f \sin{\delta} }{m v_x} + v_y \\
        A_{22} = -\frac{ C_r + C_f \cos{\delta} }{m v_x}\\        
        A_{23} = -\frac{ C_f l_f \cos{\delta} - C_r l_r }{ m v_x } - v_x \\ 
        A_{32} = -\frac{ C_f l_f \cos{\delta} - l_r C_r }{ I v_x } \\   
        A_{33} = -\frac{ C_f l_f^2 \cos{\delta} + l_r^2 C_r }{ I v_x }  \\        
		\end{array} \ .
		\end{equation}

    
    \end{subequations}    

\noindent Note that, for a easier comprenhension $C_i(\alpha_i)$ is denoted by $C_i$ being $i = f,r$.

\section{Online T-LPV-MPC using Zonotopes}

In this section, we present the zonotope-tube-based LPV-MPC (ZT-LPV-MPC) scheme to significantly reduce the computational effort in RMPC techniques for autonomous driving (see Figure \ref{fig:Robust_MPC_Kinf_escheme}).
The main purpose of this strategy is to achieve robust stability in the presence of modeling errors and bounded exogenous disturbances. 

%
%

\noindent In following subsections, the proposed ZT-LPV-MPC strategy is explained step by step for a correct comprehension.
First of all, a polytopic state feedback controller is computed offline using an H$_{\infty}$-LMI based problem. Then, in an online way, the state feedback gain is computed as a linear function of the scheduling vector $\zeta$ (Section \ref{sec:local_controller}). 
In Section \ref{sec:terminal_set}, the terminal robust invariant set and the terminal cost computations are presented to guarantee asymptotic stability of deterministic MPC.
Afterwards, in Section \ref{sec:online_reachable_Sets}, the online reachable set computation is presented
Finally, in Section \ref{sec:MPC}, the T-MPC problem is presented where the input and state constraints are updated defining an adaptive and less conservative tube.
Hereafter, the introduced scheme will be explained in detail.

\subsection{\textbf{Local controller design}} 
\label{sec:local_controller}
In this section, the offline design and online computation of the state feedback LPV controller is addressed.
We aim to design a controller to reduce the mismatch between the states of system \eqref{eq:NL-syst} and the nominal state vectors \eqref{eq:nominal-model} even under the presence of exhogenous disturbances or model uncertainty.
\noindent In the most recent literature, the Linear Quadratic Regulator (LQR) control strategy is one of the most used techniques when dealing with determining a local control structure for robustifying the MPC strategy \citep{Gonzalez_R_2011, Sakhdari_B_2017, Darup_M_2018}. 

\noindent However, when dealing with systems subject to external disturbances, the $LQR$ technique becomes less efficient against such system variations than $H_{\infty}$ strategy, that seems more interesting for the application. \\


\subsubsection{Offline design}
\noindent In this work, a polytopic LPV $H_{\infty}$ controller is designed by means of minimizing the infinity norm of the transfer function between the disturbance signal and the control variables.

\noindent In this case, the LPV representation in \eqref{eq:linear-syst-formulation} is transformed into a polytopic LPV representation for control design purposes where the scheduling vector $\zeta$ ranges now over a fixed polytope $\Theta = \{ \zeta \in \mathbb{R}^{n_{\zeta}} : H_{\zeta} \zeta \le b_{\zeta} \}$ being $n_{\zeta}$ the number of scheduling variables. Then, the polytopic representation is formulated as
\begin{equation}
\label{eq:H_inf_sys}
	\begin{aligned}
		& \dot{x} = A_{\zeta} x + B u + E d \\
		& z = C x + D_1 u + D_2 d  
	\end{aligned} \ ,
\end{equation}
where $A_{\zeta}$ is given by
\begin{equation}
    A_{\zeta} = \sum_{i=1}^{N} \mu_i (\zeta) A_i \ ,
\end{equation}
where $A_i$ represents the system dynamics at each one of the vertexes of the polytope $\Theta$. $N$ represents the number of vertexes of polytope $\Theta$ and is equal to $2^{n_{\zeta}}$. $\mu_i (\zeta)$ is the linear membership function defined by
\begin{equation}
\label{eq:RobustMPC_weighting_function}
    \mu_i (\zeta) = \prod_{j=1}^{n_{\zeta}} \xi_{ij}(\eta_{0}^{j}, \eta_{1}^{j}) \ , \ \ \ \ \ \  \forall i= 1,...,N \ , \
\end{equation}
with
\begin{equation}
\label{eq:RobustMPC_distance_functions}
\begin{aligned}
    & \eta_{0}^{j} = \frac{ \overline{\zeta_{j}} - \zeta_{j}(k)  }{\overline{\zeta_j} - \underline{\zeta_j}} \\
    & \eta_{1}^{j} = 1 - \eta_{0}^{j}   \ , 
\end{aligned}
\end{equation}
where each scheduling variable $\zeta_j$ is known and varies in a defined interval $\zeta_j \in \left[ \underline{\zeta_j}, \overline{\zeta_j} \right] \in \Theta$, $n_{\zeta}$ is the number of scheduling variables and $\xi_{ij}(\cdot)$ corresponds with the function that performs the $N$ possible combinations.
In addition, next conditions must be satisfied
\begin{equation}
	\sum\limits_{i = 1}^N {\mu_i (\zeta)}  = 1,\;\;\mu_i (\zeta) \ge 0, \;\forall \zeta \in \Theta \ .
\end{equation}

\noindent Matrix $B$ in \eqref{eq:H_inf_sys} is an instantiation of $B_{\zeta}$ in \eqref{eq:LPV_B} at $\delta = 0$ and $C_f$ at a particular constant value.
$E$ is the disturbance input matrix, $d$ represents the exogenous disturbance vector and its product $Ed$ is always contained in $W$. 
$z$ represents the controlled variables vector and $C$, $D_1$ and $D_2$ are tuning matrices of appropriate dimensions.

\noindent From the polytopic LPV system \eqref{eq:H_inf_sys} and considering the state feedback control law $u = K_{\zeta}^{\infty} x$, we can formulate the transfer function from $d$ to $z$ as
\begin{equation}
	G_{zd} =  (C + D_1 K_{\zeta})(s I - (A_{\zeta} + B K_{\zeta}^{\infty}))^{-1} E + D_2 \ . \
\end{equation}

\noindent Hence, the proposed problem consists on finding a polytopic state feedback gain $K_{\zeta}$ such that
\begin{equation}
	\norm{G_{zd}}_{\infty} \le \gamma \ ,
\end{equation}
holds for the attenuation scalar $\gamma \in \mathbb{R}$. 
To find the solution, we solve the $H_{\infty}$ problem in continuous time via LMIs using the polytopic approach as suggested in \cite{Yu_and_Duan} given by

    \begin{equation}
    \label{eq:Hinf-LMI}
    \begin{aligned}
    & \underset{X, W_i}{\text{min}}
    && \gamma      \\
    & \text{s.t.}\\    
    &&& \left[\begin{array}{ccc}
     S &  E  & X C^T + W_i^T D_2^T \\
              *     & -\gamma I &  D_1^T  \\
              *     &    *   &  -\gamma I
               \end{array}\right] \le 0 \\
    &&& S = A_i X + B W_i + X A_i^T + W_i B^T \\
    &&& \forall i = 1,...,N \ , \
    \end{aligned}
    \end{equation}
    
\noindent being the solutions $X=P^{-1} \in \mathbb{R}^{n \times n}$ and $W_i = K P^{-1} \in \mathbb{R}^{m \times n}$ where $P \in \mathbb{R}^{n \times n} > 0$ represents the common Lyapunov matrix for the polytopic LPV system.
Then, the resulting vertices of the new polytopic controller are obtained by $K_i = W_i X^{-1} \in \mathbb{R}^{m \times n}$. \\

\begin{remark}
	In H$_{\infty}$ control we have even more degrees of freedom to include additional performance weights and better attenuate unknown inputs (disturbance and noise).
\end{remark}


\subsubsection{Online computation}

%
%
%
%
%
%

At each control iteration $k$, the state feedback LPV control gain $K_{\zeta}^{\infty}$ is updated based on the current value of the scheduling vector $\zeta$.
To do so, a linear combination of polytopic controllers, i.e. the set of $K_i$, is computed as 
\begin{equation}
    \label{eq:RobustMPC_interpolation_K}
    K_{\zeta}^{\infty} = \sum_{i=1}^{N} \mu_i (\zeta) K_i \ .
\end{equation}

\subsection{\textbf{Terminal Robust Invariant Set \& Cost}}
\label{sec:terminal_set}
%
%
%

A commonly used approach to guarantee asymptotic stability of deterministic MPC consists in incorporating both a terminal cost, $P$, and a terminal constraint set, $\chi_{f}$.
In this section, we propose an offline method to compute both $P$ and $\chi_{f}$. 
Thus, the closed-loop system convergence to the origin is ensured if 
\begin{itemize}
	\item $Q = Q^T \ge 0$, $R = R^T > 0$ and $P > 0$
	\item The sets $X$, $\chi_f$ and $U$ are zonotopes containing the origin
	\item The terminal cost is a Lyapunov function in $\chi_f$
	\item $\chi_f$ is the minimal robust positively invariant (mRPI) set, $\chi_f \subseteq X$.
\end{itemize} 

On one hand, the computation of $P$ is carried out by solving the continuous LMI-based H$_{\infty}$ problem \eqref{eq:Hinf-LMI}. Furthermore, a polytopic robust controller is found. The optimal problem solutions, i.e. $X$ and $W_i$, are used to calculate the controllers at the vertices of the polytope as $K_i = W_i X^{-1}$. Note that the Lyapunov function in the optimization problem is found to be equal to $X^{-1}$ and will be used later in \eqref{eq:R-MPC} as $P$.

\noindent On the other hand, the terminal set $\chi_f$ will be the mRPI set if and only if it is contained in any closed RPI set and is convex and unique.
Then, the mRPI set for the stable and disturbed system \eqref{eq:error_dynamics} is computed by the following recursive procedure
\begin{equation}
    \label{eq:mRPI_Set}
    \begin{aligned}
    & \textnormal{1. Initialization: } \\
    & \ \ \ \ \ \Omega_0 = E_{k^*} \\
    & \textnormal{2. Loop: } \\
    & \ \ \ \ \ \Omega_{k+i} = \mathcal{A} (\Omega_k) \oplus W \\
    & \textnormal{3. Termination condition: } \\ 
    & \ \ \ \ \ \textnormal{stop when } \Omega_{k+1} = \Omega_{k} \textnormal{. Set } \chi_f = \Omega_{k+1} 
    \end{aligned} \ ,
\end{equation}
where $E_{k^*}$ is defined in the following and $\mathcal{A}(\cdot)$ is the set mapping defined as
\begin{equation}
	\mathcal{A}(\Omega_k) = \textnormal{Conv} \big\{ \bigcup_{i=1}^{N} (A_i + B K_i^{\infty}) \Omega_k \big\}.
\end{equation}
Note that, Conv$\{\cdot\}$ represents the convex hull and is used to compute the one-step reachable set for the polytopic system case. This allows to preserve the convexity of the resulting set within the recursive iterations.
However, this recursive approximation to compute the mRPI set is intractable and not realistic since we may need infinite iterations to reach the termination condition.
For that reason, in \citep{mRPI_LPV_uncertainties}, the authors propose an outer approximation method for computing the mRPI set with a given precision.
This approach consists on replacing the termination condition in \eqref{eq:mRPI_Set} by the condition of terminating when there exist a $k^{\dagger}$ iteration such that 
\begin{equation}
	\mathcal{A}^{k^{\dagger}}(\Omega_0) \subseteq \mathbb{A}_{p}^{n_x}(\epsilon),
\end{equation}
where $\mathbb{A}_{p}^{n_x}(\epsilon) = \{ x \in \mathbb{R}^{n_x} : \norm{x}_p \le \epsilon \}$ defines a ball of arbitrary small size.
Therefore, in such an article, it is concluded that the set $\Omega_{k^{\dagger}}$ is an outer approximation of the mRPI set $\Omega_{\infty}$ with the given precision $\mathbb{A}_{p}^{n_x}(\epsilon)$ as well as an RPI set too.

\noindent In addition, the initialization condition in \eqref{eq:mRPI_Set} is still not defined. To find $E_{k^*}$, which is an RPI set for the system \eqref{eq:error_dynamics}, it is necessary to solve the following iterative algorithm where there exist a finite $k^*$ such that the termination condition is reached
\begin{equation}
    \label{eq:Ek*}
    \centering
    \begin{aligned}
    & \textnormal{1. Loop: } \\
    & \ \ \ \ \ \mathcal{A}(E_k) = \textnormal{Conv} \big\{ \bigcup_{i=1}^{N} (A_i + B K_i^{\infty}) E_k \big\} \\
    & \ \ \ \ \ \overline{E}_{k+1} = \mathcal{A}(E_k) \oplus W \\
    & \ \ \ \ \ E_{k+1} = \textnormal{Conv} \{ \overline{E}_{k+1} \bigcup E_k \} \\
    & \textnormal{2. Termination condition: } \\
    & \ \ \ \ \ \textnormal{stop when } E_{k^*+1} = E_{k^*}
    \end{aligned} \ .
\end{equation}
    
\noindent Furthermore, given the stabilized system \eqref{eq:error_dynamics}, the initial convex set $E_0 \supseteq \Omega_{\infty}$ can be computed as 
\begin{equation}
	E_0 = \sum_{i=0}^{p^*-1} \mathcal{A}^i(B(r)) \oplus \frac{p^* \xi}{1-\xi} B(r) ,
\end{equation}
where $\xi \in (0,1)$, $p^* \in \mathbb{N}$ and $B(r) = \{ x \in \mathbb{R}^{n_x} : \norm{x}_{\infty} \le r \}$ is a box containing $W$.
Note that, we should find a proper $E_0$ such that $\mathcal{A}^k(B(r)) \subseteq \xi B(r)$ holds for $k \ge p^*$.

\subsection{\textbf{Online Reachable Sets}}
\label{sec:online_reachable_Sets}
This section addresses the reachable sets calculation also known as the one-step forward-reachable set computation using zonotopic-based representation.

\noindent A zonotope, represented as $\langle c_w, R_w \rangle $ with the center $c_w \in \mathbb{R}^n$ and the generator matrix $R_w \in \mathbb{R}^{n \times p}$, is a particular form of a polytope defined as the linear image of the unit cube \citep{Zonotopes_Ref}
\begin{equation}
	W = \langle c_w, R_w \rangle = \{ c_w+R_w x : \norm{x}_{\infty} \le 1 \} .
\end{equation}
Note that, the linear image of a zonotope $W = \langle c_w, R_w \rangle$ by a compatible matrix M is defined as
\begin{equation}
	M \circ W = M \circ \langle c_w, R_w \rangle = \langle M c_w, M R_w \rangle .
\end{equation}
Along this work, zonotopes are treated as centered zonotopes denoted by $\langle 0, R_w \rangle$. Then, the linear image is defined as
\begin{equation}
	M \circ W = \langle 0, M R_w \rangle 
\end{equation}
and the Minkowski sum of two centered zonotopes $W = \langle c_w, R_w \rangle$ and $G = \langle c_g, R_g \rangle$ is defined as
\begin{equation}
	W \oplus G = \langle 0, [R_w, R_g] \rangle.
\end{equation}

\noindent In this work, zonotopes are used to compute reachable sets and therefore, the tube to implement the robust MPC architecture.
The main reason for the use of zonotopes lies in their simplicity to operate with sets. Therefore, a set operation such as the Minkowski sum is reduced to a simple matrix addition.
Note that, the use of Minkowski sum or difference of two polytopes is costly, however, using zonotopes the computational cost is reduced allowing a fast computation of basic sets operations \citep{Zonotope_VS_Polytope}. 
These sets define the problem of finding the set of states that can be reached from a given set of states in a set of finite steps \citep{Borrelli_MPC_book} .
In this approach, the main idea of using reachability theory is to bound the maximum achievable values for the mismatch error \eqref{eq:error_dynamics} between the prediction model and the real measurements at every sampling time.
To this aim, the one-step robust reachable set from the set $\Phi$ is denoted as
\begin{equation}
\begin{aligned}
	& Reach(\Phi, W) = \{ y : \exists x \in \Phi, \exists u \in U, \exists w \in W \\  
	& \ \ \ \ \ \ \ \ \ \ \ \ \ \ \ \ \ \ \ \ \ \ \ \ \ \ s.t. \ y = (A_{\zeta} + B_{\zeta} K_{\zeta}^{\infty}) x + w \} .\end{aligned}
\end{equation}

\noindent Note that, by using zonotopic notation, the robust reachable set Reach($\Phi$, W) can be compactly written as 
\begin{equation}
	Reach(\Phi, W) = \{ ( (A_{\zeta} + B_{\zeta} K_{\zeta}^{\infty}) \circ \Phi ) \oplus W  \} .
\end{equation}
Then, denoting the first initial reachable set as a null zonotope ($\Phi_0 = \langle 0_{n \times 1}, 0_{n \times p} \rangle$) and the disturbance set as a constant predefined zonotope ($W = \langle c_w, R_w \rangle$), at every sampling time $k$ a group of reachable sets is computed by
\begin{equation}
    \label{eq:RobustMPC_Reach_Sets}
    		\begin{aligned}
	    & \Phi_{k+i+1} = (A_{\zeta_{k+i}} + B_{\zeta_{k+i}} K_{\zeta_{k+i}}^{\infty}) \Phi_{k+i} \oplus W \\                             
	    & \forall i = 0, ..., H_p
	    \end{aligned} \ ,
\end{equation}
where $H_p$ is the prediction horizon of the MPC strategy. \\
Note that, at time $k$, a number of $H_p + 1$ reachable sets are computed. 
Since the scheduling variables can be measured/estimated and computed, as the case of $\delta$, $\Phi_{k+0}$ is considered as $W$.
Then, the computation of each reachable set $\Phi_{k+i+1}$ will depend on its past realisation $\Phi_{k+i}$, the scheduling vector $\zeta_{k+i}$ for computing system matrices ($A_{\zeta_{k+i}}, B_{\zeta_{k+i}}$), the controller $K_{\zeta_{k+i}}$ and the uncertainty/disturbance set $W$.
Finally, these reachable sets are used for computing the concatenation of consecutive resulting state/input sets along the prediction horizon at each time $k$, known as tube.

\subsection{\textbf{MPC design}}
\label{sec:MPC} 
Considering the previous discussions about the terminal conditions, the local controller and the reachable sets, in this section, we focus our attention on the T-MPC implementation.
Figure \ref{fig:Robust_MPC_Kinf_escheme} shows the complete scheme used in this work.
Note that, the model predictive strategy is in charge of controlling the nominal system while the differences between the real system and the nominal one are compensated by the local controller.
Such a difference may be produced by external sources as a exogenous disturbances, unmodelled dynamics or by uncertain parameters in the nominal model.
Then, in order to guarantee robustness against all these sources, the reachable sets are used to compute the input/state space where the feasibility is ensured under the presence of the maximum disturbances considered in the design.

\begin{remark}
	Considering large disturbances acting over the vehicle implies bounding the differences between the real and the nominal system in a large zonotope which will lead to a more conservative scenario and also to the reduction of the maximum prediction horizon in the MPC design.
\end{remark} 

\begin{figure}
    	\centering
    	\includegraphics[width=85mm]{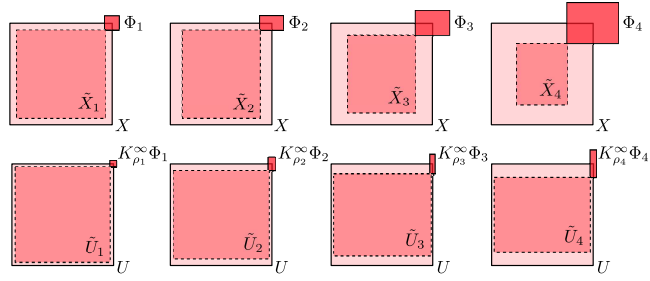}
    	\caption{Example of reachable sets ($\Phi$) growth and new MPC constraints ($\tilde{X}$ and $\tilde{U}$) evolution for a prediction horizon of four steps using a two states two inputs system. $X$ and $U$ represent the original constraints}
    \label{fig:MPC_bounds_comp}
\end{figure}

\noindent The inputs and states sets are updated at every control iteration and introduced as the new input/state constraints throughout the prediction window (see an example of a two-inputs-two-states system in Figure \ref{fig:MPC_bounds_comp}),
as follows
    \begin{equation}
    \label{eq:Input_states_sets}
    		\begin{aligned}
	    & \tilde{X}_{k+i} =  X \ominus \Phi_{k+i}  \ , \  \forall i = 0, ..., H_p , \\
	    & \tilde{U}_{k+i} =  U \ominus K_{\zeta_{k+i}}^{\infty} \Phi_{k+i} \ , \  \forall i = 0, ..., H_p-1 \ .
	    \end{aligned}
    \end{equation}
Note that, as the prediction horizon increases the possibility of reaching empty sets becomes higher resulting then in an optimal problem without solution. \\

\begin{figure}
    	\centering
		\includegraphics[width=85mm]{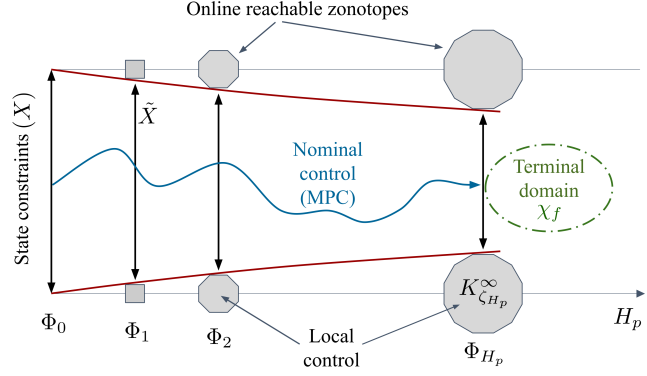}
    	\caption{ Example of the prediction stage in the ZT-LPV-MPC technique at time $k=0$. Reachable sets ($\Phi$) growth and new constraints ($\tilde{X}$) are adapted throughout this stage to guarantee robust feasability and stability}
    \label{fig:prediction_horizon}
\end{figure}  

\noindent Finally, the grouping of all the previous steps allow us to formulate the optimal problem as a quadratic optimization problem that is solved at each time $k$ (see Figure \ref{fig:prediction_horizon}) to determine the next sequence of control actions considering that the values of $x_k$ and $\tilde{u}_{k-1}$ are known
    \begin{equation}
    \label{eq:R-MPC}
    \begin{aligned}
    & \underset{\Delta U_k, X_k}{\text{min}}
    && \tilde{x}_{k+H_p}^T P \tilde{x}_{k+N} + \sum_{i=0}^{H_p-1}   ( r_{k+i} - \tilde{x}_{k+i} )^T Q ( r_{k+i} - \tilde{x}_{k+i} )  \\
    &&& +  \Delta \tilde{u}_{k+i}^T R \Delta \tilde{u}_{k+i} \\
    & \text{s.t.}\\    
    &&& \tilde{x}_{k+i+1} = \tilde{x}_{k+i} + \big( A_{\zeta_{k+i}} \tilde{x}_{k+i} + B_{\zeta_{k+i}} \tilde{u}_{k+i} \big) T_s \\
    &&& \tilde{u}_{k+i} = \tilde{u}_{k+i-1} + \Delta \tilde{u}_{k+i} \\
    &&& u_{k+i}^* \in U \ominus (K_{\zeta_{k+i}}^{\infty} \Phi_{k+i})  \\
    &&& \tilde{x}_{k+i} \in X \ominus \Phi_{k+i}  \\
    &&& \tilde{x}_{k+H_p} \in \chi_f \\
    &&& \tilde{x}_{k+0} = x_{k}   \
    \end{aligned} \ ,
    \end{equation}

\noindent where $r$ is the reference, $\tilde{x}$ is the state vector of the prediction model \eqref{eq:nominal-model}, $\tilde{u}$ is the optimal control action, $x$ is the feedback state vector from the real system, $P \in \mathbb{R}^{n \times n} > 0$ represents the terminal cost computed in Section \ref{sec:terminal_set}, $Q = Q^T \in \mathbb{R}^{n \times n} \ge 0$ and $R = R^T \in \mathbb{R}^{m \times m} \ge 0$ are the tuning matrices for the states and the variation of the control inputs, respectively.
$T_s$ represents the time period to discretize the LPV system.

\section{\textbf{Results}}

In this section, we validate the performance of the proposed ZT-LPV-MPC control scheme in a racing scenario through simulation in MATLAB.
The principal objective of the presented scheme is to follow the proposed racing-based references ensuring asymptotic stability and the highest possible level of robust performance while dealing with exogenous distrubances.

\begin{figure}
    	\centering
    	\includegraphics[width=90mm]{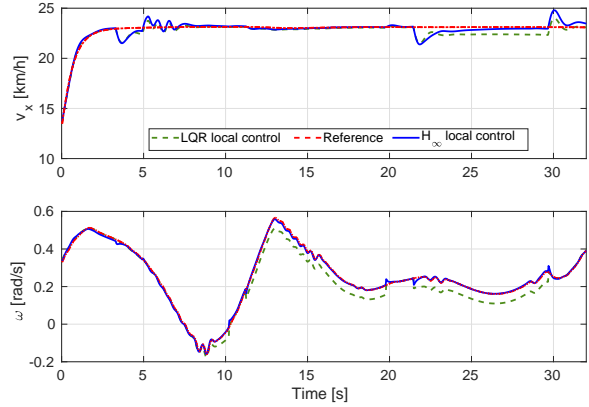}
    	\caption{ Dynamic reference tracking. Top: Longitudinal velocity reference and state $v_x$ for both compared cases. Bottom: Angular velocity reference and state $\omega$}
    \label{fig:reference_tracking}
\end{figure}

\noindent The racing references are provided by a trajectory planner \citep{Planning_Alcala_2020} and make the vehicle to perform close to its dynamic limits.
The reference vector ($r$ in Figure \ref{fig:Robust_MPC_Kinf_escheme}) is composed by two variables, the linear longitudinal speed and the angular velocity. Both are depicted as dashed lines in Figure \ref{fig:reference_tracking}.
Note that, the linear speed reference belongs to a low velocity interval, i.e. between 10 and 25 $km/h$. However, we understand a driving behaviour is closer to the limits of handling as the product between linear and angular velocities increases.
\noindent The non-linear model used for simulation is a high-fidelity bicycle-based representation of the Driverless UPC vehicle \citep{Driverless_UPC} used in the Formula Student challenge \citep{FormulaStudent} and is presented in Appendix A.
An identified tire model using the simplified Magic Formula \citep{Pacejka2005} is used for generating accurate lateral forces from front and rear slip angles.
To verify the real-time feasibility of the presented approach, we perform the simulations on a DELL inspiron 15 (Intel core i7-8550U CPU @ 1.80GHzx8). \\

\begin{figure}
    	\centering
    	\includegraphics[width=90mm]{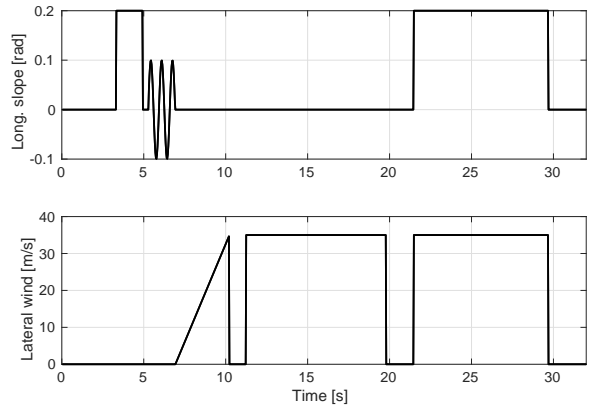}
    	\caption{ Disturbances acting on the scenario. Top: road slope profile composed by steps and sinusoid parts. Bottom: Lateral wind velocity profile in the form of steps and a ramp}
    \label{fig:disturbances}
\end{figure}

\noindent To show the effectiveness of the H$_{\infty}$-based approach presented in this work for computing the tube (Section \ref{sec:local_controller}), we perform a comparison against the LQR-based technique presented in \cite{Sakhdari_B_2017} but designed for our presented vehicle model in \eqref{eq:LPV_model}.
Hence, the comparison scenario is the same for both cases using the scheme presented in Figure \ref{fig:Robust_MPC_Kinf_escheme} where only the local controller changes for comparison purposes.
The proposed scenario consists on two disturbance sources affecting the non-linear vehicle while driving in simulation.
Such disturbance variables are chosen to be the road slope acting over the longitudinal vehicle dynamics ($\varphi$) and lateral wind affecting the lateral and angular vehicle dynamics ($F_w$) (see Figure \ref{fig:disturbances}).
These external disturbances contained in $w$ belong to the set $W = \{ w \in \mathbb{R}^{n} : H_{w} w \le b_{w} \}$ where
    \begin{equation}
        H_{w} = \left[\begin{array}{ccccc}
         	1 	& 0 & 0 & 0 & 0  	\\
           -1	& 0 & 0 & 0 & 0  	\\
         	0 	& 1 & 0 & 0 & 0 	\\
         	0 	&-1 & 0 & 0 & 0		\\
         	0 	& 0 & 1 & 0 & 0		\\
         	0 	& 0 &-1 & 0 & 0		\\
         	0 	& 0 & 0 & 1 & 0		\\
         	0 	& 0 & 0 &-1 & 0		\\
         	0 	& 0 & 0 & 0 & 1		\\
         	0 	& 0 & 0 & 0 &-1		         	 
    	\end{array} \right] , \
       	b_{w} = \left[\begin{array}{c}
         	0.074 \\
         	0.074 \\
         	0.192 \\
         	0.192 \\
         	0.105 \\
         	0.105 \\
         	0.0 \\
         	0.0 \\
         	0.0 \\
         	0.0 \\         	         	        	
    	\end{array} \right] .
    \end{equation}

\noindent The ZT-LPV-MPC uses the predicted data in the past realisation to instanciate the state space matrices at every time step within the MPC prediction stage. 
Then, the optimal control problem \eqref{eq:R-MPC} is solved at a frequency of 30 Hz using the solver GUROBI \citep{Gurobi} through YALMIP \citep{Lofberg_2004} framework and the local controller is run at a higher frequency of 200 Hz. 
The tuning parameters for the robust LPV-MPC and LPV-H$_{\infty}$ problems are listed in Table \ref{table:IJC_LPV_MPC_parameters} and \eqref{eq:IJC_H_inf_design}, respectively.

\begin{table}[!t] \label{table:IJC_LPV_MPC_parameters}
\caption{Tube-based LPV-MPC design parameters. $Q$ and $R$ matrices are normalized by dividing the respective variable by its interval to the square ($\iota^2$)}
{\begin{tabular*}{20pc}{@{\extracolsep{\fill}}llll@{}}
\hline
    Parameter & Value & Parameter & Value \\\hline
    $Q$  & 0.8*diag($\frac{0.4}{\iota_{v_x}^2}$ 0 $\frac{0.6}{\iota_{\omega}^2}$ 0 0)	& $R$ & 0.2*diag($\frac{0.5}{\iota_{\delta}^2}$  $\frac{0.5}{\iota_{a}^2}$) \\
    $\overline{\tilde{x}}$ 			& [15 1 1.4 $\infty$ $\infty$]     	& 	$\underline{\tilde{x}}$    	& [1 -1 -1.4 $-\infty$ $-\infty$]  	 \\
    $\overline{\tilde{u}}$ 			& [0.267 13]     	& 	$\underline{\tilde{u}}$    	& [-0.267 -2]  	 \\
    $\overline{\Delta \tilde{u}}$	& [0.05 0.5]  		&	$\underline{\Delta \tilde{u}}$& [-0.05 -0.5] \\
    $T_s$   	& 33 $ms$    					&   $H_p$     	& 5   \\\hline
\end{tabular*}}{}
\end{table}

\begin{subequations}
\label{eq:IJC_H_inf_design}
    \begin{equation}
     P = 10^4 \left[\begin{array}{ccccc}
   		 0.0200 &  -0.0000 &   0.0000 &  -0.0268 &  -0.0003 \\
   		-0.0000 &   0.0252 &  -0.0128 &   0.0000 &   0.0146 \\
    	 0.0000 &  -0.0128 &   0.0110 &  -0.0000 &  -0.0089 \\
   		-0.0268 &   0.0000 &  -0.0000 &   0.0376 &   0.0004 \\
   		-0.0003 &   0.0146 &  -0.0089 &   0.0004 &   1.8130 \\
         \end{array}\right] ,
   	\end{equation}
    \begin{equation}
     E = 0.3 \left[\begin{array}{ccccccc}
                 0 & 0.5 & 0 & 0 & 0 & 0 & 0  \\
                 2 & 0 & 0 & 0 & 0 & 0 & 0  \\
                 3 & 0 & 0 & 0 & 0 & 0 & 0  \\
                 0 & 0.0001 & 0 & 0 & 0 & 0 & 0  \\
                 0.01 & 0 & 0 & 0 & 0 & 0 & 0  \\
         \end{array}\right] ,
   	\end{equation}
	\begin{equation}
     C = 10^{-4} \left[\begin{array}{ccccc}
                 0.2 & 0 & 0 & 0 & 0   \\
                 0 & 0.2 & 0 & 0 & 0   \\
                 0 & 0 & 0.2 & 0 & 0 	\\
                 0 & 0 & 0 & 0.2 & 0 	\\
                 0 & 0 & 0 & 0 & 0.1  	\\
                 0 & 0 & 0 & 0 & 0		\\
                 0 & 0 & 0 & 0 & 0  	\\
         \end{array}\right] ,
	\end{equation}
	\begin{equation}
     D_1 = 10^{-4} \left[\begin{array}{ccccccc}
                 0 & 0 & -0.2 	& 0 	& 0 	& 0 	& 0  \\
                 0 & 0 & 0 		& -0.2 	& 0 	& 0 	& 0  \\
                 0 & 0 & 0 		& 0 	& -0.2 	& 0 	& 0  \\
                 0 & 0 & 0 		& 0 	& 0 	& -0.2 	& 0  \\
                 0 & 0 & 0 		& 0 	& 0 	& 0 	& -0.1  \\
                 0 & 0 & 0 		& 0 	& 0 	& 0 	& 0  \\
                 0 & 0 & 0 		& 0 	& 0 	& 0 	& 0  
         \end{array}\right] ,
	\end{equation}
	\begin{equation}
     D_2 = 10^{-3} \left[\begin{array}{ccc}
                 0 & 0  \\
                 0 & 0  \\
                 0 & 0  \\
                 0 & 0  \\
                 0 & 0  \\
                 0 & 0  \\
                 0.15 & 0  \\
                 0 & 0.15  \\
         \end{array}\right] ,
	\end{equation}
    \begin{equation}
        H_{\zeta} = \left[\begin{array}{ccccc}
         	1 	& 0 & 0 & 0 & 0  	\\
           -1	& 0 & 0 & 0 & 0  	\\
         	0 	& 1 & 0 & 0 & 0 	\\
         	0 	&-1 & 0 & 0 & 0		\\
         	0 	& 0 & 1 & 0 & 0		\\
         	0 	& 0 &-1 & 0 & 0		\\
         	0 	& 0 & 0 & 1 & 0		\\
         	0 	& 0 & 0 &-1 & 0		\\
         	0 	& 0 & 0 & 0 & 1		\\
         	0 	& 0 & 0 & 0 &-1		         	 
    	\end{array} \right] , \
       	b_{\zeta} = \left[\begin{array}{c}
         	10 \\
         	-0.5 \\
         	0.6 \\
         	0.6 \\
         	1.0 \\
         	1.0 \\
         	1000.0 \\
         	0.0 \\
         	3.14 \\
         	3.14 \\         	         	        	
    	\end{array} \right] .
    \end{equation}	
\end{subequations}

\noindent The reference tracking results are depicted in Figure \ref{fig:reference_tracking}. It can be seen the significant improvement of the presented scheme with respect to the ZT-LPV-MPC using LQR controller as the corrective error approach ('LQR local control' in figures).
Furthermore, the disturbance rejection has enhanced using a local controller whose design has been based on minimizing the infinity norm instead of the 2-norm as the case of LQR approach.
However, note that using a $H_{\infty}$ design may produce troubles in the closed-loop response because of the large gains that are obtained and hence, a meticulous tuning is needed.
\begin{figure}
    	\centering
    	\includegraphics[width=90mm]{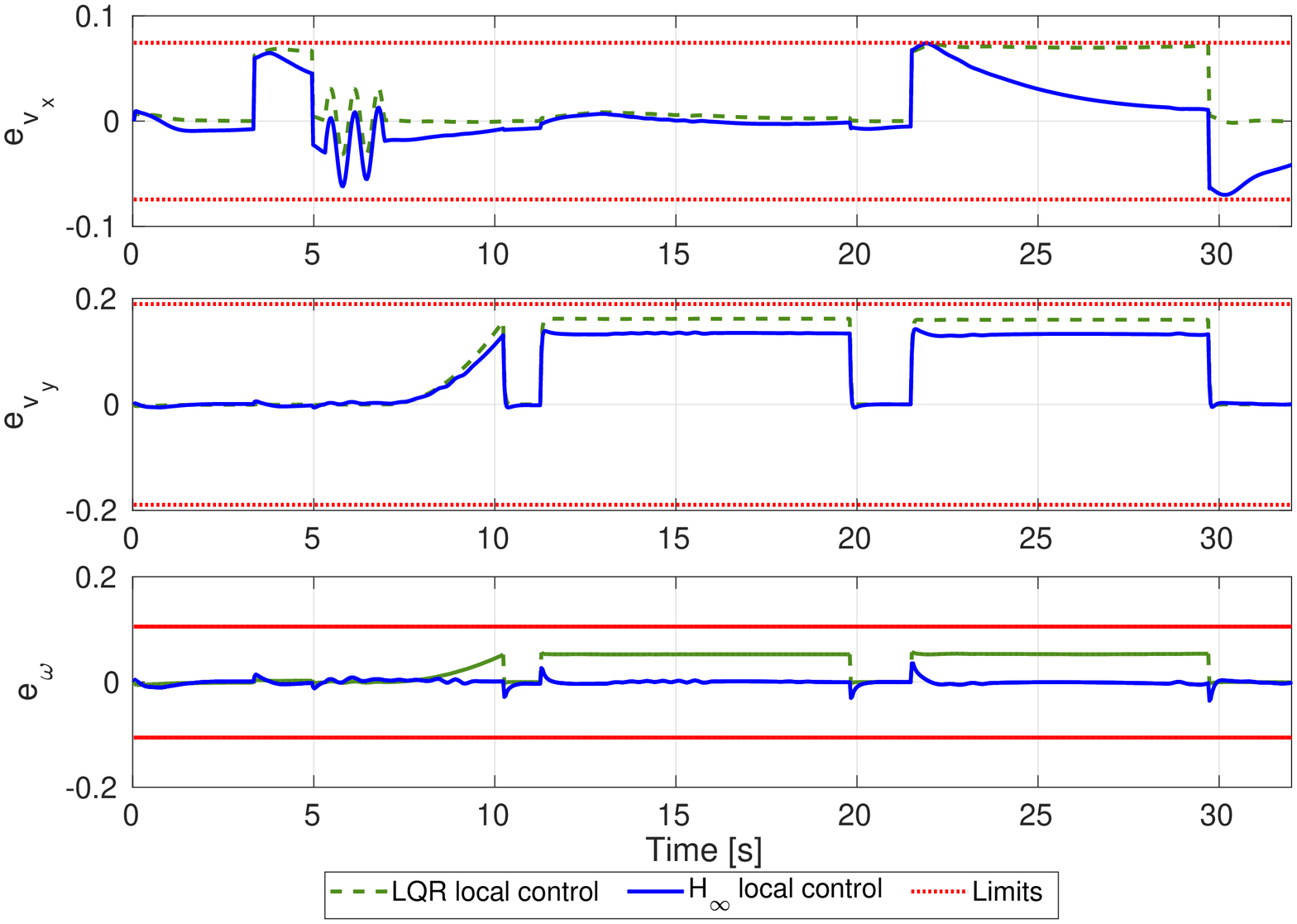}
    	\caption{ Mismatch between real and nominal states. $e_{v_x}$ represents the error in the longitudinal behaviour, $e_{v_y}$ the error in the lateral behaviour and $e_{\omega}$ represents the error for the angular behaviour. Dotted red lines represent the maximal bounds for each one of th errors defining then the set $W$}
    \label{fig:errors}
\end{figure} 

\noindent In Figure \ref{fig:errors}, the errors or mismatch between the predicted state and the measured state are presented. Note that, such a vector of errors correspond with the vector entering the state feedback local controller ($e$ in Figure \ref{fig:Robust_MPC_Kinf_escheme}).
It can be appreciated the better performance of the strategy presented in this work being able to reject most of the error produced by the uncertainty and the applyied exogenous disturbances.
\begin{figure}
    	\centering
    	\includegraphics[width=90mm]{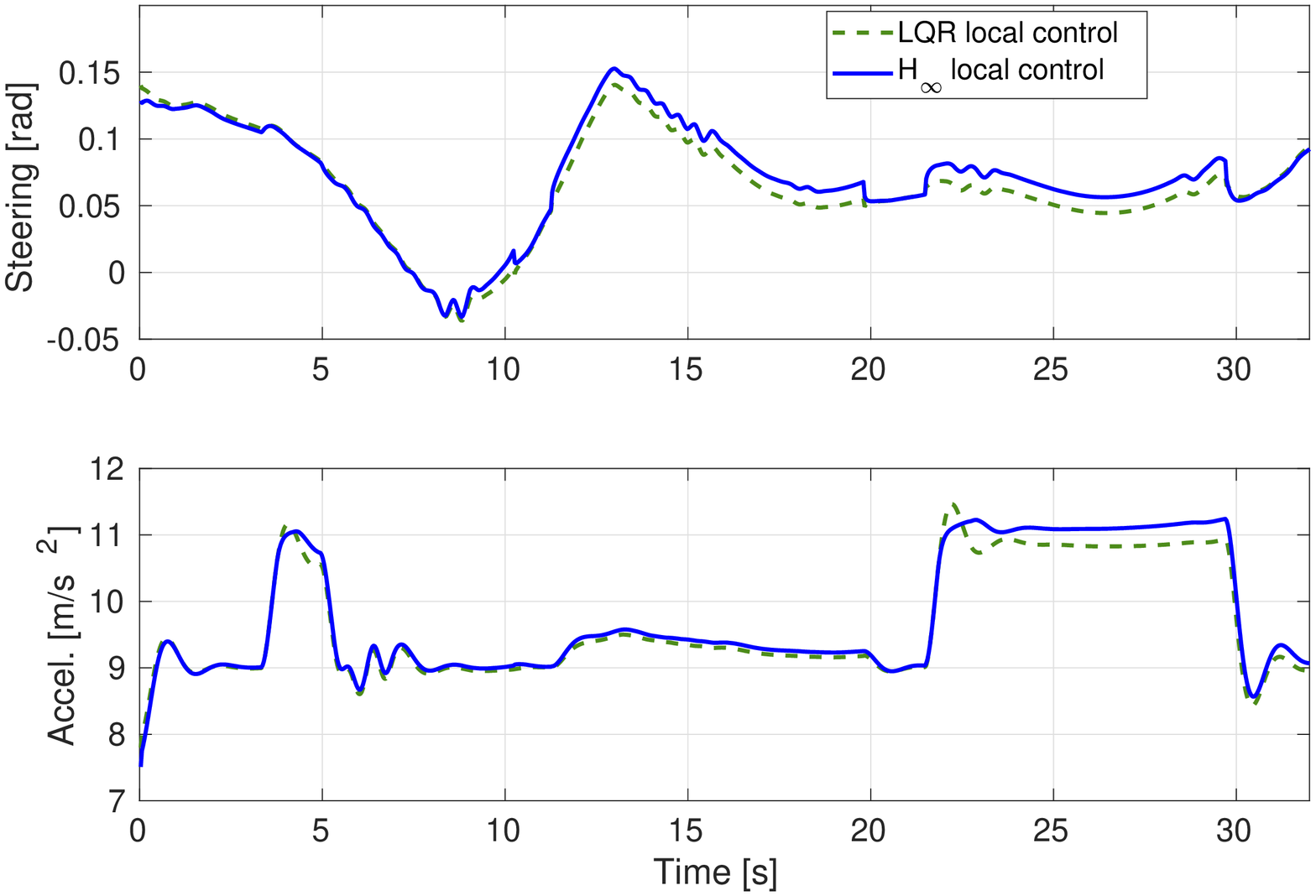}
    	\caption{ Control actions applyied to the simulated vehicle ($u$ in Figure \ref{fig:Robust_MPC_Kinf_escheme})}
    \label{fig:control_inputs}
\end{figure} 

\begin{figure}
    	\centering
    	\includegraphics[width=90mm]{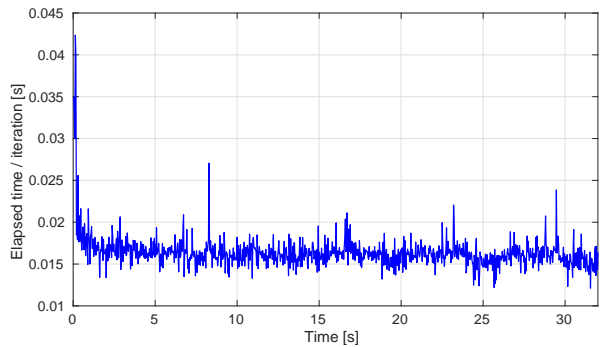}
    	\caption{ Elapsed time per iteration during the simulation. The mean time is 0.0164 s}
    \label{fig:elapsed_time}
\end{figure}

\noindent Figure \ref{fig:control_inputs} shows the control actions applyied during the simulation test.
Figure \ref{fig:elapsed_time} shows the elapsed time per iteration of the ZT-LPV-MPC strategy where the mean elapsed time per iteration is 16.4 ms using a prediction horizon of 5 steps.

\noindent In addition, in Table \ref{table:Time_comparison_zonotope_polytope}, we performe an elapsed time comparison between polytope-based and zonotope-based operations for computing the tube in a particular time instant. In this, we show a computational improvement when using zonotopes of around 285 times faster than using standard polytope formulation. 

\begin{table}[!t]
\caption{Reachable set computational time comparison for a sequence of length $H_p$\label{table:Time_comparison_zonotope_polytope}}
{\begin{tabular*}{10pc}{@{\extracolsep{\fill}}ll@{}}
\hline
    Approach & Mean Computation Time \\
    \hline
    Polytopic	& 4 ms		\\
    Zonotopic	& 0.014 ms	\\
    \hline
\end{tabular*}}{}
\end{table}

\begin{figure}
    	\centering
    	\includegraphics[width=90mm]{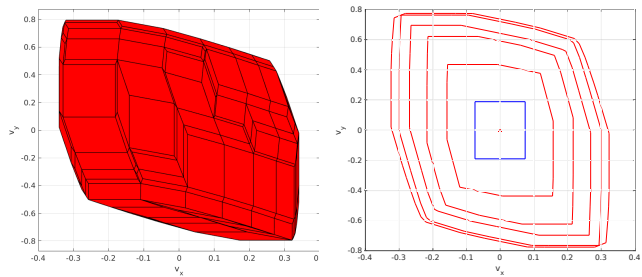}
    	\caption{ Reachable sets representation for $v_x$ and $v_y$. Left side: external view of the reachable set computed for the last prediction in the LPV-MPC. Right side: cross-section view of the evolution of each one of the reachable sets computed online at a particular time instant}
    \label{fig:ReachSets_zonotope_polytope}
\end{figure}

\noindent Figure \ref{fig:ReachSets_zonotope_polytope} shows both an external set view showing the exact realisation of the reachable set in the last iteration of the prediction horizon (left side) and a cross-section view of each one of the reachable sets during the prediction horizon (right side).
It is important to highlight that the exact propagation of the reachable sets using zonotopes is made online at a very low computational cost (see Table \ref{table:Time_comparison_zonotope_polytope}).
Thus, we prove the fast tube computation using zonotope theory. \\ 

\noindent Finally, a quantitative comparison is made using the normalized root mean squared error (NRMSE) as performance measurement (see Table \ref{table:NRMSE}).
These results highlight the conclusive improvement of the proposed approach, improving up to thirty times the angular velocity tracking error with respect to the compared strategy in the proposed disturbed scenario.

\begin{table}[!t]
\caption{Quantitative results for the tracking variables errors. These are the difference with respect to their respective reference (see Figure \ref{fig:reference_tracking})\label{table:NRMSE}}
{\begin{tabular*}{20pc}{@{\extracolsep{\fill}}lll@{}}
\hline
    Approach & RMSE $v_x$ &  RMSE $\omega$ \\
    \hline
    ZT-LPV-MPC with LQR local control 	& 4.3846 $10^{-4}$ 	& 0.0249		\\
    ZT-LPV-MPC with H$_{\infty}$ local control	& 3.4227 $10^{-4}$ 	& 8.0762 $10^{-4}$\\
    \hline
\end{tabular*}}{}
\end{table}

\section{\textbf{Conclusion}}

A Zonotope-Tube-based Linear Parameter Varying MPC (ZT-LPV-MPC) scheme for autonomous driving is proposed for handling fast and disturbed scenarios.
The proposed approach uses an LPV representation of the vehicle to predict the future behaviour and design a gain-scheduling LPV-H$_\infty$ controller to ensure fast convergence of the mismatch between real and predicted states on disturbed scenarios. Besides, the computational cost is further improved at this point with respect to other alternatives in the literature (see Table \ref{table:T_MPC_Classification}).

\noindent Using reachability theory using zonotopes, the MPC changes online its state and input constraints to ensure robust stability under exhogenous disturbances.
In addition, we prove the fast and accurate tube computation using zonotopes instead of polytopes.

\noindent Finally, we test the presented scheme and compare the H$_{\infty}$-based local controller performance against the LQR design for the local controller.
The framework was tested on a fast disturbed scenario, demonstrating significant performance improvements in disturbance rejection and computation time (achieving a mean elapsed time of 16.4 ms) compared to the current state of the art results achieved in this field.
Further research will focus on extending the vehicle model as well as implementing and validating the proposed strategy in a experimental platform.

\section*{Acknowledgments}
This work has been funded by the Spanish Ministry of Economy and Competitiveness (MINECO) and FEDER through the projects SCAV (ref. DPI2017-88403-R) and HARCRICS (ref. DPI2014-58104-R). 
The author is supported by a FI AGAUR grant (ref 2017 FI B00433). \\

\section{Appendix}

\subsection{Vehicle model for simulation}

For simulation purposes we use a higher fidelity vehicle model. Unlike the model used for control design, this considers a more precise tire model, i.e. the Pacejka "Magic Formula" tire model where the parameters $b$, $c$ and $d$ define the shape of the semi-empirical curve. Also, a more accurate computation of the tire slip angles is given. 
  	
\noindent Notice that, variables $\varphi$ and $v_w$ are exogenous disturbances and represent the longitudinal road slope and the lateral wind velocity.  	
Furthermore, $C_{dAl}$ is the product of drag coefficient and vehicle lateral cross sectional area.
Parameters $d_f$,$d_r$,$c_f$,$c_r$,$b_f$ and $b_r$ are the simplified Pacejka model constants. 

\noindent All the vehicle parameters are properly defined in Table \ref{table:vehicle_parameters}.

	\begin{equation}
   		\label{eq:NL_model_for_simulation}
    		\begin{aligned}
       	& \dot v_x = a_r + \frac{- F_{yf} \sin{\delta} - F_{df}}{m} + \omega v_y - g \sin{\varphi}  \\
        	& \dot v_y =  \frac{ F_{yf} \cos{\delta} + F_{yr} - F_{w}}{m} - \omega v_x \\
		& \dot \omega = \frac{ F_{yf} l_f \cos{\delta} - F_{yr} l_r - F_{w}(l_f - l_r)}{I} \\
    		& \alpha_f = \delta - \tan^{-1} \left(  \frac{v_y}{v_x} - \frac{l_f \omega}{v_x} \right) \\
    		& \alpha_r = - \tan^{-1}  \left(  \frac{v_y}{v_x} + \frac{l_r \omega}{v_x} \right) \\    		      
        	& F_{yf} = d_f \sin{ (c_f \tan^{-1} (b_f \alpha_f)) } \\
		& F_{yr} = d_r \sin{ (c_r \tan^{-1} (b_r \alpha_r)) } \\        
    		& F_{df} = \mu m g + \frac{1}{2} \rho C_{dAf} v_x^2  \\   		        		
    		& F_{w} = \frac{1}{2} \rho C_{dAl} v_w^2  \ . \ 
    		\end{aligned}
	\end{equation}

\begin{table}[!t]
\caption{Dynamic model parameters of the Driverless UPC Car\label{table:vehicle_parameters}}
{\begin{tabular*}{20pc}{@{\extracolsep{\fill}}llll@{}}
\hline
    Parameter & Value & Parameter & Value \\\hline
    $l_f$     & 0.902  $m$   &  $l_r$   & 0.638  $m$  \\
    $m$       & 196    $kg$  &  $I$     & 93 $kg$ $m^2$  \\
    $d_f$     & 8.255        & $c_f$    & 1.6 \\
    $b_f$     & 6.1     		& $\mu$ 	   & 1.4  \\ 
    $d_r$     & 8.255    	& $c_r$    & 1.6 \\
    $b_r$     & 6.1     		& $\rho$   & 1.225 $kg$ $m^3$ \\    
    $C_{dAf}$ & 1.64     	& $g$      & 9.81 $\frac{m}{s^2}$ \\  
    $C_{dAl}$ & 1.82         & $w$      & 1.45 $m$ \\
    $d$       & 2.3   $m$    &  \\ \hline
\end{tabular*}}{}
\end{table}

\subsection{Tire stiffness LPV model}
 
\noindent The Pacejka tire equations in \eqref{eq:NL_model_for_simulation} for front and rear wheels are reformulated in a LPV representation for a proper introduction in the final LPV vehicle model \eqref{eq:LPV_model}.
Hence, starting from previous data representing the dynamics of the tires obtained by means of experimental tests, a least-squares algorithm is used to find two polynomials fitting the experimental tire data as

\begin{equation}
	\label{eq:polynomial_pacejka}
     	F_{y}(\alpha) = p_1 \alpha^n + p_2 \alpha^{n-1} + ... + p_n \alpha + p_{n+1} \ , \ 
\end{equation}

\noindent where $p$ constants are the estimated coefficients that define the particular model structure and $n$ represents the order of the corresponding polynomial.

\noindent Once the polynomial is adjusted, the embedding approach of the non-linearities inside a varying parameter has to be used in order to obtain its LPV representation. Then, the following formulation is proposed

\begin{equation}
	\label{eq:LPV_pacejka}
     	F_{y} = C (\alpha) \ \alpha \ , \ 
\end{equation}
where
\begin{equation}
	\label{eq:LPV_stiffness_coef}
     	C (\alpha) = p_1 \alpha^{n-1} + p_2 \alpha^{n-2} + ... + p_n + p_{n+1}/(\alpha+\epsilon)
\end{equation}

\noindent is known as the tire stiffness coefficient and $\epsilon$ is a very small constant. 
Note that, as $\alpha$ becomes close to zero in \eqref{eq:LPV_stiffness_coef}, $C(\alpha)$ grows exponentially. To avoid this behaviour, a saturation is added in the small interval $\alpha \in [0, 0.0075]$ such that $C(\alpha)$ value stay at $4 \times 10^4$.
Table \ref{table:Polynomial_parameters} shows the coefficients used in \eqref{eq:LPV_stiffness_coef}.
      
\begin{table}[htbp]\label{table:Polynomial_parameters}
\caption{Polynomial parameters of \eqref{eq:LPV_stiffness_coef} for the front and rear tires (upper indexes f and r)}
{\begin{tabular*}{20pc}{@{\extracolsep{\fill}}llll@{}}
\hline
    Parameter & Value & Parameter & Value \\\hline
    $n$        & 4              &  $\epsilon$   & $10^{-4}$   \\  
    $p_1^f$    & -2.167 $\times 10^6$  &  $p_2^f$      & 1.284 $\times 10^6$  \\
    $p_3^f$    & -0.288 $\times 10^6$  &  $p_4^f$      & 0.029 $\times 10^6$  \\
    $p_5^f$    & 15.038         &               &  \\  
      
    $p_1^r$    & -2.130 $\times 10^6$  &  $p_2^r$      & 1.198 $\times 10^6$  \\
    $p_3^r$    & -0.252 $\times 10^6$  &  $p_4^r$      & 0.024 $\times 10^6$  \\
    $p_5^r$    & 14.551         &               &  \\ \hline
\end{tabular*}}{}
\end{table}

\end{document}